\documentclass[a4paper,10pt,pra,superscriptaddress,showpacs,twocolumn]{revtex4}
\usepackage[dvips]{graphics,color}
\usepackage{amsmath}
\usepackage{amsfonts}
\usepackage{amssymb}
\usepackage{amstext}
\usepackage[sort&compress]{natbib}
\usepackage[colorlinks]{hyperref}

\begin{document}

\title{Entanglement and local information access for graph states}

\author{Damian Markham }
\affiliation{Department of Physics, Graduate School of Science,
  University of Tokyo, Tokyo 113-0033, Japan}

\author{Akimasa Miyake }
\affiliation{\mbox{Institute for Theoretical Physics, University of Innsbruck,
Technikerstra{\ss}e 25, A-6020 Innsbruck, Austria} \\
\mbox{Institute for Quantum Optics and Quantum Information,
Austrian Academy of Sciences, Innsbruck, Austria}}

\author{Shashank Virmani }
\affiliation{Optics Section, Blackett Laboratory \& Institute
for Mathematical Sciences, Imperial College, London SW7 2AZ, United Kingdom}

\date{March 27, 2007}


%

\begin{abstract}
We exactly evaluate a number of multipartite entanglement measures
for a class of graph states, including $d$-dimensional cluster
states ($d = 1,2,3$), the Greenberger-Horne-Zeilinger states, and
some related mixed states. The entanglement measures that we
consider are continuous, `distance from separable states' measures,
including the relative entropy, the so-called geometric measure, and
robustness of entanglement. We also show that for our class of graph
states these entanglement values give an operational interpretation
as the maximal number of graph states distinguishable by local
operations and classical communication (LOCC), as well as supplying
a tight bound on the fixed letter classical capacity under LOCC
decoding.

\end{abstract}
\maketitle

\section{Introduction}

The understanding and quantification of entanglement can be said to
be one of the most fundamental problems in quantum information
\cite{Plenio05}. Entanglement measures often have operational
meanings. The distillable entanglement, for example, is the
asymptotic number of Bell pairs extractable by local operation and
classical communication (LOCC). Entanglement measures can also be
used to classify quantum resources, such as the necessary condition
presented in \cite{Vandennest06} for universal resources in one-way
quantum computation \cite{Raussendorf01}. Theoretical knowledge of
entanglement values for interesting states may also enable us to
estimate those of experimentally prepared states only via
measurements of linear witness operators \cite{Guehne06}. However,
apart from bipartite scenarios, the calculation of truly {\it
multipartite} entanglement measures is generally considered to be
formidable even for pure states (cf. \cite{Ishizaka05}).

We will primarily be interested in a set of simple multi-qubit
entangled states known as ``graph states''\cite{Hein06,Hein04}, or
stabilizer states (up to local unitaries), which have proven useful
in a variety of quantum information tasks. They include the
Greenberger-Horne-Zeilinger (GHZ) state, cluster states (a universal
resource for one-way quantum computing \cite{Raussendorf01}), and
Calderbank-Shor-Steane (CSS) error correction codeword states.
Graph states themselves can be seen as algorithmic specific resources
in the framework of one-way computing \cite{Raussendorf01}, due
to their common simple prescription to prepare.
Closely related weighted graph states have recently found use in
approximating ground states for strongly-interacting spin
Hamiltonians \cite{AndersS06}. Furthermore, small graph states, in
particular cluster states, are a current topic in the laboratory
\cite{Mandel03} and have been used for one-way quantum computation
\cite{Walther05}.

One of the main purposes of this paper is to exactly evaluate
continuous, distance-like multipartite entanglement measures for
graph states, on the way giving them an operational interpretation.
Our idea is to utilize the connection between these widely-studied
entanglement measures and LOCC state discrimination (see e.g.
\cite{Hayashi05} and refs. therein). This will allow us to give
upper and lower bounds to the entanglement of {\it any} graph state
by using a simple graphical interpretation of these states. In
several interesting cases these upper and lower bounds match, hence
giving the exact entanglement values. It is remarkable that our
argument does not require any difficult calculations despite the
generally troublesome optimization that is usually involved in
calculating the entanglement measures that we consider. Our method
is advantageous as not only does its graphical nature mean that it
is often easy to obtain the answer (namely the entanglement values
of interest), but it also provides intuition about the nature of
multipartite entanglement, for example, how the amount of
multipartite entanglement is related to the way Bell pairs are
stored in it.

We begin by giving some background definitions and results in
section \ref{sectn: Background}. We then present the calculation of
entanglement and LOCC discrimination protocols for sets of pure
graph states in \ref{sectn: Ent and disc}. In section \ref{sectn:
mixed case} this is extended to several related mixed states. These
results are given an operational interpretation in section
\ref{sectn: ChannelCap} where we use previous results to give tight
fixed letter channel capacities for LOCC decoding. We finish with
conclusions.

\section{Graph states and entanglement measures} \label{sectn: Background}

Graph states $|G_{k_1 \ldots k_n}\rangle$ of $n$ qubits can be
described pictorially by a graph $G$ of $n$ vertices, with $n$
binary indices $(k_1, \ldots, k_n)$ such that $k_i ={0,1}$
\cite{note_graph}. Let us denote the Pauli matrices at the $i$-th
qubit by $X_i$, $Y_i$, $Z_i$ with the identity $\openone_i$. The
vertices of $G$ represent qubits, each of which is initially
prepared in the $(-1)^{k_i}$ eigenstate of $X_i$, i.e.
$\frac{1}{\sqrt{2}}(|0\rangle + (-1)^{k_i} |1\rangle)$. The graph
states are then defined by performing 2-qubit Control-Z operations
(${\rm CZ}_{ij}= {\rm diag}(1,1,1,-1)$ in the $Z$ basis), e.g. via
an Ising interaction, on all pairs of qubits joined by the edges
${\rm Ed}$ of $G$.
It can be shown that the $2^n$ graph states $\{|G_{k_1\ldots
k_n}\rangle \}$ are the joint eigenstates of the $n$ independent
commuting operators (called stabilizer generators):
\begin{eqnarray}
\label{eqn: DEF Graph generators}
K_i := X_i \bigotimes_{(i,j)\in
{\rm Ed}} Z_j \qquad i=1,\ldots, n
\end{eqnarray}
such that the graph states satisfy the eigenequations
\begin{eqnarray}
K_i |G_{k_1 \ldots k_i \ldots k_n}\rangle = (-1)^{k_i} |G_{k_1
\ldots k_i \ldots k_n}\rangle ,
\end{eqnarray}
i.e. the index $k_i$ gives the stabilizer eigenvalue $(-1)^{k_i}$
for $K_i$. Hence if we can measure $K_i$, we can determine $k_i$. We
will later utilize this feature to design protocols for
discriminating sets of these states using LOCC. Note that for a
given $G$, the graph states $\{|G_{k_1 \ldots k_n}\rangle\}$
construct a complete orthonormal basis of dimension $D_H =2^n$. They
are local unitarily equivalent as $|G_{k_1\ldots k_n}\rangle =
\prod_{i=1}^{n}Z_{i}^{k_i} |G_{0\ldots 0}\rangle$, hence they all
have equal entanglement.

Our methods can be used to derive additive bounds on entanglement
values for any graph state. However most of the graphs that we will
consider explicitly in this paper are examples of ``two-colourable''
\cite{Aschauer05} graph states. These are defined as graph states
where it is possible to assign one of two colours (say Amber and
Blue) to each qubit, such that no two qubits connected by an edge
have the same colour. Many of the graph states that have found
significance in quantum information are of the two-colourable kind -
e.g. GHZ states, CSS codeword states, and cluster states are all
two-colourable.

Among the graph states that we consider, we are able to derive {\it
exact} entanglement values for $d$D cluster states (i.e. the
$d$-Dimensional cubic grid graphs of Figs.~\ref{fig: linear} and
\ref{fig: Cluster even}), the GHZ states (i.e. the tree graph of
Fig.~\ref{fig: GHZ} with one centre vertex and $n-1$ vertex
`leaves'), and the Steane $[[7,1,3]]$ codeword state. To our
knowledge, the entanglement values for such graph states have been
calculated only for the discrete Schmidt measure (i.e. the minimum
number of terms required in a product state expansion of the state)
in Ref.~\cite{Hein04} (cf. \cite{Fattal04}). A merit of continuous
entanglement measures is that their values are stable under small
deviation in the state space.

The entanglement measures that we consider in this paper are defined
as follows. For a state $\rho$, the {\it relative entropy of
entanglement} \cite{Vedral98} is defined as
\begin{eqnarray}
E_R(\rho) =
\min_{\omega \in{\rm SEP}} {\rm tr}\rho (\log_2 \rho - \log_2
\omega),
\end{eqnarray}
where the minimum is taken over all fully separable mixed states
$\omega$. The {\it global robustness of entanglement} \cite{Vidal99}
is defined as
\begin{eqnarray}
R(\rho) = \min_{\omega} t
\end{eqnarray}
such that there exists a state $\omega$ such that $(\rho + t
\omega)/({1+t})$ is separable. For convenience we also define an
extension of the {\it geometric measure} \cite{Shimony95} as
\begin{eqnarray}
E_g(\rho) = \min_{\omega \in {\rm SEP}} -\log_2
({\rm tr}(\rho \omega))
\end{eqnarray}
(note that $E_g$ is an entanglement monotone only for pure states
$\rho$).

In Ref.~\cite{Hayashi05} it has been shown that the maximum number
$N$ of pure states in the set $\{|\psi_i\rangle|i=1,\ldots,N\}$,
that can be discriminated perfectly by LOCC (in fact by separable
\cite{LOCCdiscrefs1} measurements), is bounded hierarchically (cf.
\cite{Wei04}) by the amount of entanglement they contain:
\begin{eqnarray}
\label{eqn: bound N}
N \leq  { D_H \over {\overline{1+R(|\psi_i\rangle)}}} &\leq&
{D_H \over{\overline{2^{E_R(|\psi_i\rangle)}}}}  \leq
{D_H \over{\overline{2^{E_g(|\psi_i\rangle)}}}},
\end{eqnarray}
where $D_H = 2^n$ is the total dimension of the Hilbert space, and
$\overline{x_i}=\frac{1}{N}\sum_{i=1}^N x_i$ denotes the
``average''. In fact it can be shown, using a local symmetry
argument (``twirling''), that for {\it all} pure stabilizer states
the two rightmost inequalities of Eq.~(\ref{eqn: bound N}) collapse
to equalities \cite{note_twirling}. However, although these
inequalities give powerful insight into the relationship between
LOCC state discrimination and the quantification of entanglement,
computing any of the measures in Eq.~(\ref{eqn: bound N}) is usually
extremely difficult. Hence it is not clear to what extent these
relationships will enable quantitative progress. The aim of our
paper is to show that in fact, for large classes of graph states,
these inequalities may be exploited to derive strong explicit bounds
that in many cases are exact. As many of the graph states that we
consider have played a diverse and important role in the literature,
it is likely that our computations will prove useful in
understanding the role of these entanglement measures in multiparty
quantum information scenarios.

\section{LOCC Discrimination and entanglement of graph
states} \label{sectn: Ent and disc}
We are now ready to apply Eq.~(\ref{eqn: bound N}) to graph states
with a given graph $G$. In order for the states to perfectly
distinguishable, they must be orthogonal. Hence we consider the
perfect LOCC discrimination of a subset of the complete orthonormal
basis $\{|G_{k_1 k_2 \ldots k_n}\rangle\}$. Since all states in the
set have equal entanglement, i.e. $E(|G_{k_1\ldots k_n}\rangle) =
E(|G_{0\ldots 0}\rangle) \; \forall k_i$, the average in
Eq.~(\ref{eqn: bound N}) can be replaced by $E(|G_{0\ldots
0}\rangle)$.

We will evaluate the hierarchy of inequalities in Eq.~(\ref{eqn:
bound N}) from above and below in terms of graph problems. Adopting
the notation of equation ~(\ref{eqn: bound N}), for a given choice
of graph $G$ let $N$ denote the largest number of graph states
associated with $G$ that may be perfectly discriminated using LOCC.
In subsection (A) below we will derive a simple lower bound on $N$
by finding sets of graph states that {\it can} by construction be
explicitly distinguished by LOCC. In subsection (B) we will provide
an upper bound on the rightmost term of equation (\ref{eqn: bound
N}) based upon a very elementary analysis of the bipartite
entanglement present across certain splittings of the graph states.
We will then show that these two bounds meet for many sets of
two-colourable graph states mentioned above, hence giving both their
exact entanglement values as well as the maximal possible $N$. Our
methods are also applicable to certain mixed states, which we will
discuss in a later section.

\begin{table}[t]
\begin{tabular}{c||c|c|c}
\;&\; $2^{\max m_c}$ & $2^{n-\max m_p}$ & $E_g, \; E_R, \; \log_2(1+R)$\\
\hline {\rm $d$D cluster} & $2^{\lceil \frac{n}{2} \rceil}$ &
$2^{n - \lfloor\frac{n}{2}\rfloor} = 2^{\lceil\frac{n}{2}\rceil}$ &
$\lfloor \frac{n}{2}\rfloor$ \\
{\rm GHZ} & $2^{n-1}$ & $2^{n-1}$ & 1  \\
{\rm ring} & $2^{\lfloor \frac{n}{2}\rfloor}$ &
$2^{n-\lfloor\frac{n}{2} \rfloor} = 2^{\lceil \frac{n}{2}\rceil}$
&
$\lfloor \frac{n}{2}\rfloor \leq E \leq \lceil \frac{n}{2}\rceil$\\
Steane code & $2^4$ & $2^{7-3}$& $3$
\end{tabular}
\caption{Summary of the lower and upper bounds, and entanglement values
\cite{note_odd}. Information encoded on these states, can be decoded
by LOCC with capacity $C=n-E$.}
\label{table:bounds}
\end{table}

\subsection{Lower ``colouring'' bound}
A lower bound for $N$ can be given by maximizing the number $m_c$ of
stabilizer generators $\{K_i\}$ that can be determined
simultaneously in a {\it single} setting of LOCC measurements. If we
can evaluate $m_c$ eigenvalues of stabilizer generators by LOCC, we
know that we can discriminate deterministically at least $2^{m_c}$
states, by picking only one state from each subspace determined by
the $m_c$ eigenvalues. For example, by finding the eigenvalues of
$\{K_i\}_{i=1}^{m_c}$, we can discriminate the set of $2^{m_c}$
states $\{|G_{k_1 k_2 \ldots k_{m_c}0\ldots 0}\rangle\}$. Therefore,
we have a lower bound as
\begin{equation}
2^{m_c}\leq 2^{\max m_c} \leq N. \label{eqn: lower}
\end{equation}

One approach is to identify a set of $|A|$ qubits that are not
connected to each other and are thus coloured by a single colour
(we call these the Amber qubits). By locally
measuring $X$ on all the Amber qubits, and locally measuring $Z$ on
all the others, one can use Eq.~(\ref{eqn: DEF Graph generators})
and classical communication to determine the eigenvalues $\{k_i |
i\in A\}$ corresponding to the subset of generators $\{K_i|i\in
A\}$. We can see this as follows. Without loss of generality we
label the amber qubits ${i=1,\ldots, |A|}$. By Eq.~(\ref{eqn: DEF
Graph generators}) we see that the Amber generators
$\{K_i\}_{i=1}^{|A|}$ have only $\openone$ or $X$ on the first $|A|$
qubits, and then either $Z$ or $\openone$ on the remaining ones. Hence
the measurement of the generators from the set $\{K_i\}_{i=1}^{|A|}$
can be simulated by LOCC measurement by locally measuring $X$ on all
the Amber qubits, and locally measuring $Z$ on all the others, and
then communicating the outcomes. In order to get as tight a bound as
possible we would like to have $m_c=|A|$ as large as possible. Hence
Eq.~(\ref{eqn: lower}) translates into a bound
\begin{eqnarray}
2^{|A|} \leq N,
\end{eqnarray}
where $|A|$ is the (maximum) number of mutually disconnected qubits,
i.e., the (maximum) number of vertices which can be coloured
by the same single colour.
In graph theoretic terminology the set of vertices that achieves
such a maximum is known as the {\it maximum independent set}.

In the case of two colourable graphs this analysis can be
particularly simple. For a given 2-colouring of the graph one can
set the colour with the larger number of vertices to be Amber and
the other to be Blue, $B$, i.e., set $m_c = |A| \geq |B|$. We
illustrate this colouring strategy in Fig.~\ref{fig:
linear}-\ref{fig: Cluster even}. In those figures we readily find that
the optimal lower bound that can be obtained using this approach is
given for GHZ states by setting all the leaf vertices as Amber, i.e.
$\max \; m_c = n-1$, whereas for $d$D cluster states we find that
$\max \;m_c = \lceil \frac{n}{2}\rceil$. The best lower bounds for
ring states (closed 1D chain graph) and Steane codeword states can
be shown in a similar way - the results are summarized in
Table~\ref{table:bounds}.

\begin{figure}[t]
\resizebox{7cm}{!}{\includegraphics{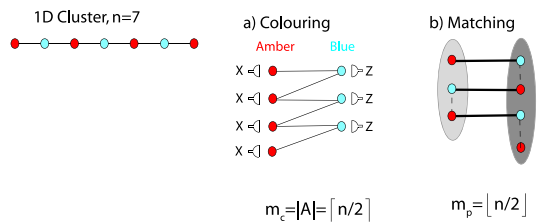}} \caption{ We illustrate the example of a 1D Cluster state
with an odd number of qubits. a) {\it Colouring}: Without loss of generality, we can always redraw the
graph in two columns, of $|A|$ Amber and $|B|$ Blue qubits, where the number of Amber qubits $|A|>|B|$.
We then measure the Amber qubits in the X basis, and the Blue in the Z basis. There is no entanglement
within the Amber column, or within the Blue column, but our protocol works for arbitrary entanglement
between the two columns (and indeed, any entanglement between the Blue qubits for non 2-colourable
states). b) {\it Matching}: We group the qubits into two parts (the shaded regions), local unitaries
w.r.t. this partition, i.e., local Control-Z, get rid of the extra entanglement (in the dotted lines),
leaving only maximally entangled pairs (the bold lines). This matching and colouring is easily extended
to give bounds for the ring states in Table~\ref{table:bounds}. }
\label{fig: linear}
\end{figure}

One may enquire whether these lower bounds may be improved upon by taking into account the fact that
different graphs may correspond to physical states that are equivalent under local unitary
transformations (i.e. with the same entanglement properties). Indeed, as we show in Fig.~\ref{fig:
GHZ}, for the GHZ state there are two different graphs - the fully connected one, and the usual `tree'
graph. The fully connected graph has a maximum independent set of $1$ vertex, whereas the `tree' gives
a maximum independent set of $n-1$. Hence by considering different graphs corresponding to the same
graph state (up to local unitary transformations) one may obtain vastly improved lower bounds. However,
it turns out that for the examples in this paper this freedom does not lead to better lower bounds than
the ones that we present in Table~\ref{table:bounds}. This includes in particular the ring graph state
for which there exists a gap between the lower and upper bounds. We will however use this freedom in
the next section (via the method of {\it Local Complementation}) where it will enable us to improve the
upper bounds that we will obtain for certain types of graph.

Note that the LOCC identification of stabilizer elements is also an
important primitive in entanglement distillation
\cite{Aschauer05,Miyake05}. The only real differences being the fact
that in distillation protocols stabilizer eigenvalues are determined
{\it indirectly} (i.e., their parities) in order not to destroy
entanglement, and furthermore {\it all} stabilizer eigenvalues are
evaluated to get a specific target pure graph state from an ensemble
of identical noisy copies.

\begin{figure}[t]
\resizebox{8cm}{!}{\includegraphics{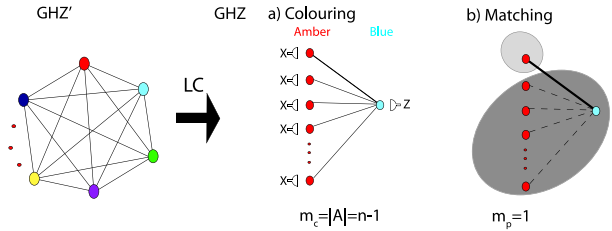}} \caption{ Colouring and Matching for a GHZ state. Here we see
an extreme example where we can go from one graph to another by local unitary transformation of the
state (i.e. not changing the entanglement properties), improving dramatically the colouring bound of
Eq.~(\ref{eqn: lower}). By performing local complementation (LC) (see section \ref{subsection:
matching}) on any one of the qubits we can go from a $n$-colourable fully connected state GHZ' to the
standard GHZ state increasing the maximum number of independent vertices from $1$ to $n-1$. From there
it is easy to see how to achieve the optimum colouring and matching. Graph states in the literature,
however, are often already in the optimum form, as is the case for all our other examples.} \label{fig:
GHZ}
\end{figure}

\subsection{Upper ``matching'' bound} \label{subsection: matching}

We will obtain an upper bound to the rightmost term of
Eq.~(\ref{eqn: bound N}), by weakening the constraint of
full-separability. If we define $E_{g_{bi}}$ as the geometric
measure with respect to some {\it bipartition}, we have that $E_g
\geq E_{g_{bi}}$ since the set of fully separable states is a subset
of the bipartite separable states. Hence,
\begin{eqnarray}
{D_H \over{ \overline{2^{E_g(|\psi_i\rangle)}}}} =
2^{n-E_g(|G_{00\ldots 0}\rangle)} \leq 2^{n-
\max E_{g_{bi}}(|G_{00\ldots 0}\rangle)} .
\end{eqnarray}
Our strategy is to try to find a suitable bipartition across which
the entanglement is as large as possible. Once we have specified a
bipartition we can readily calculate the entanglement across it by
several methods (cf. Refs.~\cite{Hein04,Audenaert05}). However, we
will pose this question as a graph ``matching'' problem in order to
gain intuition into when the bounds can be tight.

In particular, we will consider transforming a graph by bipartite
LOCC into another graph made only of $m_p$ disjoint ``matched'' Bell
pairs (we may extend these ideas also to other tree-type units).
Since local unitaries leave bipartite entanglement unchanged, the
entanglement is then simply $E_{g_{bi}}=m_p$.

\begin{figure}[t]
\resizebox{9cm}{!}{\includegraphics{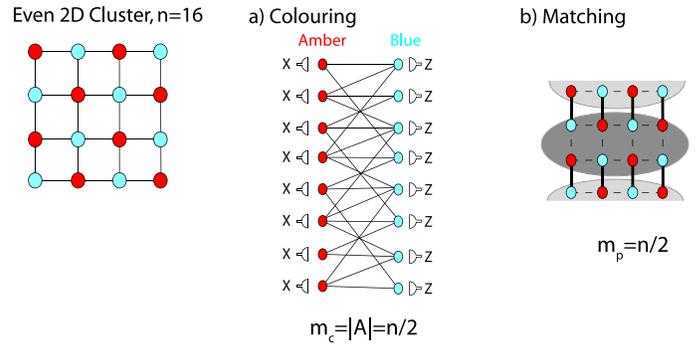}}
\caption{ Colouring and Matching for a $4 \times 4$ 2D Cluster state.}
\label{fig: Cluster even}
\end{figure}

The simplest cases arise when {\em local} applications of Control-Z
can be used to erase edges within each partition, as we illustrate
in figures (\ref{fig: linear}-\ref{fig: Cluster even}). This simple
approach is often sufficient to match the upper bounds we obtained in
section (A). For example, this approach works for even cluster
states (cf. Figs.~\ref{fig: linear}-\ref{fig: Cluster even}), even
ring states, GHZ states, and the Steane code.

\begin{figure}[t]
\resizebox{8.5cm}{!}{\includegraphics{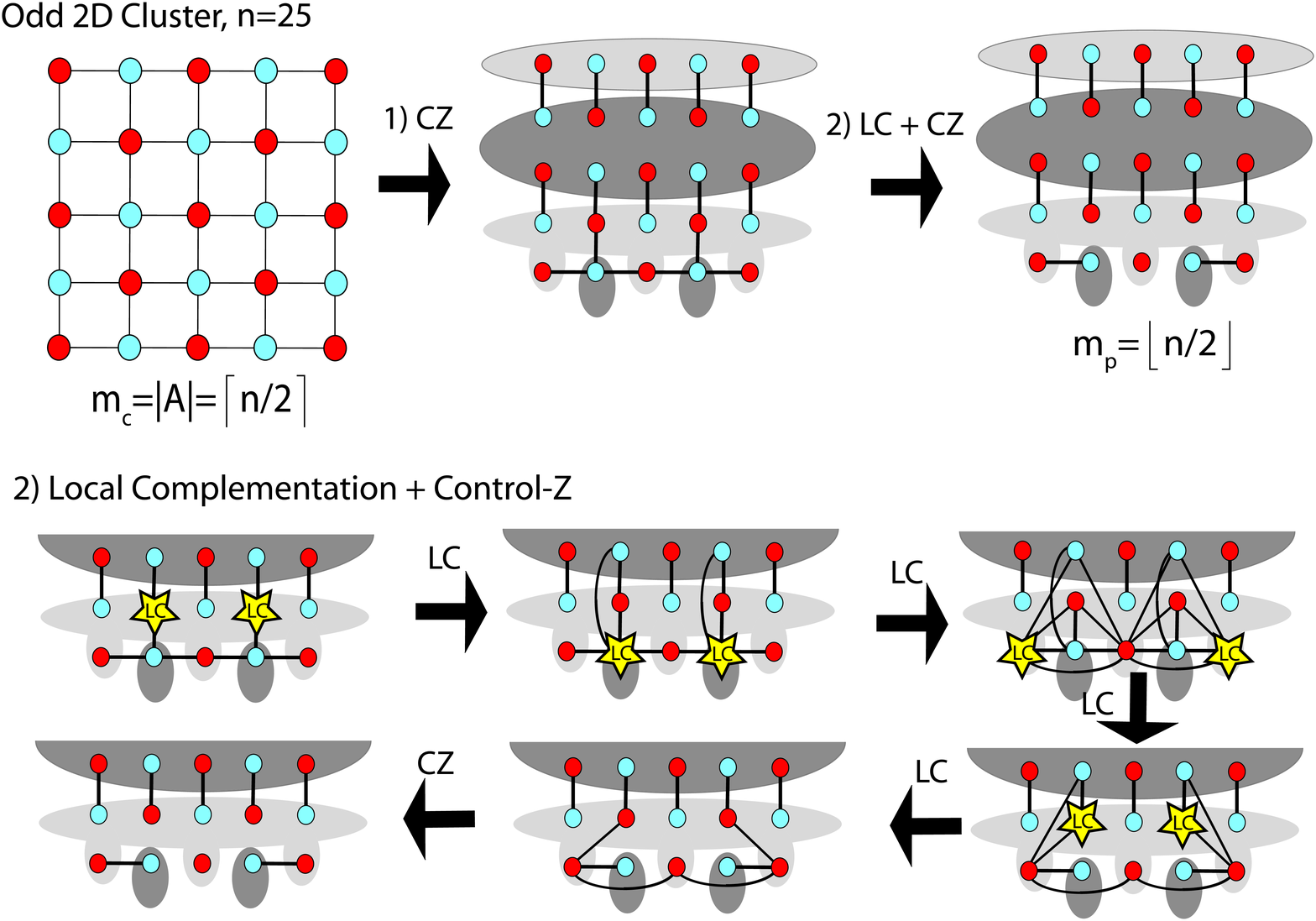}} \caption{
Matching for the $5 \times 5$ 2D cluster state with a proof of
feasibility by LOCC. The qubits are partitioned into two parts, the
light and dark shaded regions. The matching is achieved by bipartite
LOCC in two steps: 1) {\it Local} control-Z erase edges within each
partition, leaving a bunch of Bell pairs as well as a ``comb'' graph
in the odd case. 2) A comb graph in the bottom odd-numbered row can
be transformed into Bell pairs, by local complementation (LC)
unitaries indicated by the yellow stars, followed by a final CZ.
This method can be extended without to much difficulty to prove the
matching for 3D cluster states and the Steane code state.}
\label{fig: Cluster odd}
\end{figure}

For more complicated graphs such an elementary approach does not always work, and so we must utilize
the so-called {\it local complementation} (LC) of graph states \cite{Hein06}, which corresponds to a
multi-local unitary operation $V_i$ on the $i$-th qubit and its neighbors, defined as $ V_i =
\sqrt{K_i} = \exp \left(-\frac{i\pi}{4}X_i \right) \prod_{(i,j)\in{\rm Ed}} \exp \left(\frac{i\pi}{4}
Z_j \right)$. LC centred on a qubit $i$ is visualized readily as a transformation of the subgraph of
$i$-th qubit's neighbours, such that an edge between two neighbours of $i$ is deleted if the two
neighbours are themselves connected, or an edge is added otherwise. The use of LC to transform the odd
2D cluster state into a bunch of Bell pairs is illustrated in Fig.~\ref{fig: Cluster odd}. The
technique of LC often enables us to match the lower bounds derived using the approach of section (A).
For example, although we do not present further details, it turns out that LC can easily be used to
derive the optimum number of Bell pairs for all 1D, 2D or 3D cluster states, including those with an
odd total number of qubits. In the case of ring graphs with an odd number of vertices, examples of
non-two-colourable graphs, LC does not enable us to exactly match the bounds of section (A), but
nevertheless enables us to achieve quite close bounds on the exact value (see Table 1). In all of the
cases discussed above, except for the odd ring states, this bound matches the lower colouring bound and
we have equivalence for all of the measures in equation (\ref{eqn: bound N}) (cf. Table 1).

It is interesting that for important states such as cluster states,
all of these multipartite entanglement values coincide with
bipartite entanglement ones (as well as the Schmidt measure values
in Ref.~\cite{Hein04}). Note however that it is never possible to
transform these graph states into a disjoint set of Bell pairs if we
work under the original fine-grained multiparty LOCC, so the
entanglement values represent more than just bipartite entanglement.

In summary, for {\it all} graphs, we have
\begin{equation}
n-|A| \geq \log_2 (1+R(|G\rangle)) = E_R(|G\rangle) = E_g(|G\rangle)
\geq E_{g_{bi}}(|G\rangle),
\end{equation}
where $|A|$ is the maximal possible number of mutually disconnected
qubits in the graph (as mentioned in section \ref{sectn: Background}
the central equalities follow from a local symmetry argument
\cite{note_twirling}). For two colourable graphs, we can set
$n-|A|=|B|$. Since the RHS can always be calculated for some
bipartition, we can always calculate simple upper and lower bounds
to all these entanglement quantities. However, we have seen that if
it is possible to use the graph transformations above with an
appropriate bipartition to get $ \max m_p=|B|$ Bell pairs of
bipartite entanglement, as in our examples, then all entanglement
measures are equal to $|B|$. We were able to show such equality
above and calculate the entanglement for 1D, 2D, 3D Cluster states,
GHZ states, and the Steane code state. Good bounds are given for the
ring state. These results are summarised in table \ref{table:bounds}
and the proofs illustrated in Figs.~\ref{fig: linear}-\ref{fig:
Cluster odd} (the ring state, 3D cluster and the Steane code states
proofs are simple adaptations of the figures and so not shown).

Furthermore, for these states the entanglement quantities are
additive, since bipartite pure state entanglement is additive for
these measures. This will prove useful in section \ref{sectn:
ChannelCap} where we consider discrimination in an asymptotic
setting for calculating classical channel capacities. This
additivity also has implications for investigations of asymptotic
distillation of multipartite quantum states, where regularized
(asymptotic) measures such as the relative entropy of entanglement
can play an important role, see e.g. Ref.~\cite{Fortescue}.

\section{Classes of Mixed States} \label{sectn: mixed case}

The above methods also allow us to compute the entanglement of
certain mixtures of graph states. The reason for this is that in
obtaining the exact entanglement values for the graph states
discussed above, we are actually also able to derive an explicit
form for the `closest' separable state in the various entanglement
measures. If two or more pure graph states share the same `closest'
separable state, then one can exploit the fact that mixtures of
these states will also share the same `closest' state, and this
enables us to compute the entanglement of such mixtures. In more
detail, the argument proceeds by the following steps:

(1) {\it Each joint eigenspace of the Amber generators $\{K_i|i\in
A\}$ can be spanned by product states.} Suppose the set of Amber
generators $\{K_i|i\in A\}$ is simultaneously determined as
discussed previously. The outcomes of these measurements determine
one of the joint eigenspaces of the operators $\{K_i|i\in A\}$. Each
such eigenspace has dimension $2^{n - |A|}$, and more crucially, it
can be shown that each of these eigenspaces can be spanned by
product states. This can be seen as follows - if we pick a new set
of stabilizer generators $\{K_i |i\in A\} \cup \{Z_i | i\notin A\}$,
then these operators are all mutually commuting, and it is not too
difficult to verify that they are the stabilizers of product states.
As this new set of stabilizers contains the Amber generators,
$\{K_i|i\in A\}$, this means that it must be possible to span the
joint eigenspaces of the $\{K_i|i\in A\}$ by product states.

(2) {\it Consider any graph for which
$E_R(|G\rangle)=E_g(|G\rangle)=|B|$. In such cases, for each graph
state $|G\rangle$ selected from a given eigenspace $S_A$ of the
operators $\{K_i|i\in A\}$, a closest separable state for the
relative entropy and geometric measures can be taken to be the equal
mixture of all product states spanning $S_A$.} The relative entropy
between a given pure graph state $|G\rangle$ and the equal mixture
of all pure states spanning $S_A$ can easily be calculated as:
\begin{equation}
-\log_2 ({1 \over 2^{n-|A|}}) = |B|
\end{equation}
As the conditions of the statement assert that this achieves
$E_R(|G\rangle)$, we know that in these cases the equal mixture of
states spanning $S_A$ must provide an optimal separable state.
Similar arguments apply for the geometric measure.

(3) {\it Consider any graph for which
$E_R(|G\rangle)=E_g(|G\rangle)=|B|$. In such cases, for any mixture
of graph states selected from a given eigenspace $S_A$ of the
operators $\{K_i|i\in A\}$, a closest separable state for the
relative entropy and geometric measures can be taken to be the equal
mixture of all product states spanning $S_A$.} This follows
straightforwardly from the previous statement, and the fact that if
the same separable state $\omega$ is optimal for two states
$\rho_1,\rho_2$, then it is also optimal for mixtures $p_1 \rho_1 +
(1-p_1)\rho_2$ (this in turn follows from the convexity of both the
geometric measure functional and the relative entropy).

Thus, consider any mixture $\rho$ of graph states
$|G_{\vec{k}}\rangle$ from $S_A$, i.e.,
\begin{eqnarray} \label{eqn: mix k}
\rho = \sum_{\vec{k}} \lambda_{\vec{k}} |G_{\vec{k}}\rangle\langle
G_{\vec{k}}|,
\end{eqnarray}
where the $\lambda_{\vec{k}}$ are the eigenvalues of $\rho$, and are
nonzero only for indices $\vec{k}$ such that the $\{k_i |i \in A\}$
take constant (but otherwise arbitrary) values. The relative entropy
and geometric measures can be computed for such mixed states as
\begin{align}
\begin{split}
E_R(\rho) &= |B|- S(\rho), \\
E_g(\rho) &= |B|,
\label{eq: E mixed}
\end{split}
\end{align}
where $S(\rho)= -{\rm tr}\rho \log_2 \rho$.
Note that these results include any binary mixture of Bell basis
states for the 2-qubit case, and these expressions are additive for
tensor products of these states.

(4) {\it Computing the robustness $R$ for such mixed states.} For
such mixed states one can also derive the robustness of entanglement
as
\begin{eqnarray}
R(\rho) = 2^{|B|}\max \lambda_{\vec{k}} -1.
\label{eqn: rob mixed}
\end{eqnarray}
This is based on the following lower bound for the robustness of a
state given the robustness and weight of one state in its convex
decomposition, the derivation of which is shown in footnote
\cite{proof_robustness}:
\begin{equation}
\label{eq:lower robust}
R(\rho) \geq p_0 (1+R(\rho_0)) - 1 .
\end{equation}
It is easy to show that this lower bound is achieved
when $\rho$ is a mixture of graph states from the same joint
eigenspace $S_A$, by admixing in the minimal $\omega$ that turns
$\rho$ into an equal mixture of pure states spanning $S_A$, thereby
giving equation (\ref{eqn: rob mixed}).

We can also extend the above analysis to consider some mixtures of
pure graph states corresponding to different graphs. Provided that
(a) the states being mixed are defined for the same number $n$ of
qubits, (b) they all have entanglement
$E_R(|G\rangle)=E_g(|G\rangle)=|B|$, (c) the pure graph states being
mixed are taken from the eigenspace $S_A$, and (d) the generators
for the Amber qubits do not change, then Eqs.~(\ref{eq: E
mixed}) still stand. For example, take any two colourable graph $G$
used in the previous sections, and apply the local complementation
operation to any of the Amber qubits to generate graph $G'$. Clearly
two (potentially non-orthogonal) states $|G_{\vec{k}}\rangle$
and $|G'_{\vec{k}}\rangle$ have the same entanglement since
the transformation is local.
Also, the generators $\{K_i| i \in A\}$ remain unchanged
since the edges to neighbors of Amber qubits are unaffected. Thus,
following the same logic as above, our entanglement measures $E_R$ and
$E_g$ for the mixed state of the form
\begin{eqnarray}
\rho = u |G_{\vec{k}}\rangle\langle G_{\vec{k}}| +(1-u)
|G'_{\vec{k}}\rangle\langle G'_{\vec{k}}|,
\end{eqnarray}
are given by the same formula in Eqs.~(\ref{eq: E mixed}).
The analysis used to derive the formula for the robustness in
Eq.~(\ref{eqn: rob mixed}) does not seem to be extended to this case
straightforwardly.


%
\section{Classical capacity of quantum multiparty channels}
\label{sectn: ChannelCap}
Imagine we have encoded classical information onto multipartite
quantum states, we can ask how well we can access information in
these states locally, as an application of the preceding results. We
begin by extending Eq.~(\ref{eqn: bound N}) to the probabilistic
case. Suppose that we have been given a state from an ensemble
$\{\rho_i\}$ which we will measure with an LOCC POVM $\{M_j\}$. The
conditional probabilities of getting each measurement outcome are
given by $p(j|i):= {\rm tr} (M_j\rho_i)$. In the manner of
\cite{Hayashi05}, we bound these conditional probabilities in terms
of the entanglement of each $\rho_i$. Any POVM element in an LOCC
measurement can be written as $M_j=s_j\omega_j$, where $s_j={\rm tr}
M_j$, and $\omega_j$ is a {\it separable} normalized quantum state.
The conditional probability of {\it successful} discrimination is
hence bounded as
$p(i|i)=s_i{\rm tr}(\omega_i\rho_i) \leq s_i 2^{-E_g(\rho_i)}$ ,
where the last inequality follows as $\omega_i$ must be separable.
Due to the completeness of the POVM elements $\sum_i s_i = D_H$,
this condition can be rearranged and bounded as
\begin{eqnarray}
\sum_i p(i|i)2^{E_g(\rho_i)} \leq D_H .
\label{eqn: prob}
\end{eqnarray}
This equation can prove useful in the analysis of fixed-letter
channel capacities using graph state codewords with LOCC readout.
Following logic similar to \cite{Winter05}, suppose that Alice
transmits codewords of length $L$ formed from strings of states
$\{\rho_i\}$. If we require that the receivers must decode using
LOCC measurements with worst case conditional error bounded as
$1-p(i|i)<\epsilon$, then Eq.~(\ref{eqn: prob}) will give a bound on
the rate of the code.
If we assume that the geometric measure is additive under tensor
products of our signal states, as is the case for all the
two-colourable examples discussed in previous sections, then the
maximum number $N(L)$ of possible codewords of length $L$ must be
bounded, according to Eq.~(\ref{eqn: prob}) with $D_H = 2^{L n}$,
as:
\begin{equation}
\log_2 \{N(L)\}/ L \leq n - \overline{E_g(\rho_i)} - \log_2
\{(1-\epsilon)\}/L .
\end{equation}
In the large blocklength limit $L \rightarrow \infty$ the third term
vanishes, and this gives a bound on the capacity of:
\begin{eqnarray}
C \leq n - \overline{E_g(\rho_i)}. \label{eqn: capacity}
\end{eqnarray}
This general bound holds whenever the states $\{\rho_i\}$ in the ensemble have a geometric entanglement
that is {\it additive} (note that additivity does not always hold \cite{Werner02}, though it does for
all states considered here). This capacity bound is achievable (tight) for our examples in
Table~\ref{table:bounds} by selecting $\{\rho_i\}$ from the appropriate subspaces of graph states and
discriminating them perfectly using the colouring protocols in section \ref{sectn: Ent and disc}.

However, for pure graph states the bound of Eq.~(\ref{eqn:
capacity}) in all these examples is no better than a similar bound
by {\it bipartite} entanglement measures derived recently in
Ref.~\cite{Winter05}, since the geometric entanglement $E_g$
unfortunately reduces to bipartite entanglement. Nevertheless, our
discussion also applies to the mixed states discussed in the
previous section, and also shows that no tighter bound than
Eq.~(\ref{eqn: capacity}) can be derived using only the entanglement
properties (i.e. local unitarily invariant functions) of these
states.

\section{Conclusion}
We have introduced a simple graphical strategy, via Eq.~(\ref{eqn:
bound N}), to evaluate several distance-type multipartite
entanglement measures for graph states and have shown exact values,
seen in Table~\ref{table:bounds}, for interesting graph states such
as $1,2,$ and $3$D cluster states with an arbitrary number of
qubits. The lower and upper bounds in our evaluation can be
formulated as widely-studied graph problems (up to local unitary
equivalence of graph states). The lower bound is rephrased as the
maximum independent set problem \cite{note_minvertex} (by
associating such a set with Amber) which is known to be NP-complete
in general. On the other hand, the upper bound is formulated as the
maximum matching problem for which polynomial time algorithms exist
in general graphs (cf. Ref.~\cite{Ambainis05} for quantum
algorithms), but we have to check additionally whether all erased
edges are attributed to local operations in a given matching. Taking
advantage of existing approximate algorithms in graph theory, it may
be possible to obtain a good estimate for entanglement values for
wider graph states. It is also conceivable that Local
Complementation may be used to improve both the lower and upper
bounds in the cases where they are not tight. Although searching
over the complete orbit of a given graph under local complementation
appears to be exponentially complicated in the number of qubits
\cite{Danielson}, the results that we present in
Table~\ref{table:bounds} show that for many interesting classes of
graph state the canonical choice of graph can often already give
tight bounds.

Recently in Ref.~\cite{Vandennest07}, our results have been found to
have a direct application to the research of one-way quantum
computation, too. Since entanglement is simply consumed in the
course of one-way computation, roughly speaking, the amount of
multipartite entanglement in the initial resource states must be
high enough to be capable to carry on universal computation. It is
shown that a suitably chosen entanglement measure (like the
geometric measure we addressed) gives a necessary criterion for
universal resources. Since the geometric measure for a known
universal resource state for one-way computation, namely the 2D
cluster state, grows unboundedly with the system size $n$, any
universal resource state must have an unbounded amount of
entanglement for the geometric measure, as well. This criterion
immediately implies, for example, that the GHZ state, for which $E_g
= 1$ regardless of $n$, cannot be a universal resource.

We are grateful for very helpful discussions with M. Owari, M.
Hayashi, M. Murao, H. Briegel, M. Plenio and J. Eisert. This work
was supported by the FWF, the European Union (OLAQUI, SCALA, QAP,
QICS), JSPS, the Leverhulme Trust, and the Royal Commission for the
Exhibition of 1851.


\begin{thebibliography}{99}
\bibitem{note_graph}
Here we refer to any state $|G_{k_1 \ldots k_n}\rangle$ as a graph state associated with $G$, although
this name is sometimes reserved for $|G_{0 \ldots 0}\rangle$ in the literature.
\bibitem{note_twirling}
The twirling operation required is the averaging over the local unitary group generated by the
stabilisers corresponding to the graph state. This result will be presented in a separate paper.
\bibitem{note_odd}
For a positive integer $x$, $\lceil \frac{x}{2} \rceil$ and $\lfloor \frac{x}{2} \rfloor$ mean
$\frac{x}{2}$ if $x$ is even, and $\frac{x+1}{2}$, $\frac{x-1}{2}$ respectively if $x$ is odd.
\bibitem{proof_robustness}
The note gives a proof of Eq.~(\ref{eq:lower robust}). Given a convex decomposition of a state $\rho =
\sum_{k=0,\ldots } p_k \rho_k$, a real number $t\in\mathbb{R}^+$, and another state $\omega$ such that:
\begin{equation*}
\rho + t \omega \in SEP.
\end{equation*}
Then we see that:
\begin{eqnarray*}
 & p_0 \rho_0 + t \omega + \sum_{k>0} p_k \rho_k  \in SEP \nonumber \\
\Rightarrow & \rho_0 + \left({(t + \sum_{k>0} p_k) / p_0}\right) \tilde{\omega} \in SEP, \nonumber
\end{eqnarray*}
where $\tilde{\omega}$ is the normalised state:
\begin{equation*}
(t \omega + \sum_{k>0} p_k \rho_k)/(t + \sum_{k>0} p_k).
\end{equation*}
From these equations and the definition of the robustness we see that:
\begin{eqnarray*}
& (t + \sum_{k>0} p_k) / p_0 \geq R(\rho_0)  \\
\Rightarrow & t \geq p_0 R(\rho_0) - \sum_{k>0} p_k  \\
\Rightarrow & t \geq p_0 R(\rho_0) - (1-p_0)  \\
\Rightarrow & R(\rho) \geq p_0 R(\rho_0) - (1-p_0)  \\
\Rightarrow & R(\rho) \geq p_0 (1+R(\rho_0)) - 1.
\end{eqnarray*}
\bibitem{note_minvertex}
A solution of this problem gives the same bound by the minimum vertex covering problem which was used
in Ref.~\cite{Hein04} to bound the Schmidt measure.
\end{thebibliography}
\end{document}